\documentclass[a4paper]{mem}
\usepackage{natbib}
\usepackage{graphicx}
\usepackage[a4paper]{hyperref}
\idline{0}{0}

\begin{document}

\title{RR Lyrae Stars in the Halo: Tracers of Streams of Debris of 
Disrupted Galaxies
\thanks{Based on observations at the Observatorio Nacional de Llano del Hato, 
Venezuela}
}

\author{A. Katherina Vivas \inst{1,2} \and Robert Zinn \inst{2}\fnmsep}

%\offprints{A. K. Vivas}

\institute{Centro de Investigaciones de Astronom{\'\i}a (CIDA),
Apartado Postal 264, M\'erida 5101-A. Venezuela. \email{akvivas@cida.ve}\\
\and Yale University. Department of Astronomy. PO Box 208101. New Haven, 
CT 06511, USA
}

\abstract{
We discuss the first part of a survey for RR Lyrae variables in the
galactic halo that is being made with the 1m Schmidt telescope at the 
Venezuelan National Observatory.  So
far the survey has discovered 497 variables in 380 deg$^2$, lying from 4
to 60 kpc from the Sun.  It has detected three statistically
significant clumps of variables, which shows that outer halo does not
have smooth density contours.  One clump is located at 50 kpc from the 
galactic center,
and it is probably tidal debris from the Sagittarius dwarf spheroidal
galaxy.  A second structure, at 17 kpc from the galactic center,
appears to be due to the tidal disruption of the globular cluster Pal
5.  The third group, at 19 kpc, is not related with any
known globular cluster or dwarf galaxy.  The little that
is known about its properties is consistent with it being debris from
a disrupted galaxy.

\keywords{Surveys --
	  Stars: variables: RR Lyr --
          Galaxy: halo --
          Galaxy: structure
          }
}

\authorrunning{Vivas \& Zinn}
\titlerunning{RR Lyrae Stars in the Halo}
\maketitle

\section{Introduction}

Wide field imagers are essential tools for modern studies of the large-scale 
structure of the halo of our Galaxy since it is necessary to observe 
hundreds of square degrees of the sky. 
A second key ingredient is
to use a tracer of the halo population that
is easily separated from the numerous foreground stars belonging to
the thin and thick disk populations.
Among several other good possibilities, variable stars of the RR Lyrae type
are ideal tracers. They are low-mass evolved stars and therefore, they trace
the oldest stellar population (age $> 9$ Gyrs). They 
are well known standard candles, which make them ideal for studying
the three-dimensional structure of the halo. And finally, they are very easy to
recognize because of their large-amplitude variability ($\sim 1$ mag) and
relatively short periods.

In recent years our understanding of the structure of the halo has advanced
significantly. It is now clear that at least the outer regions of the halo, 
do not have a smooth distribution of stars. Instead, the distribution
seems to be quite clumpy \citep{new02,viv01,yan00,ive00}. The most likely
interpretation is that these sub-structures are relics of small satellite
galaxies that
have been accreted and destroyed by the tidal forces of the Milky Way. 
The stars (and globular clusters) of those disrupted satellite galaxies 
have formed a part, if not all of the halo of our Galaxy. 
This scenario is also consistent with theoretical
models of hierarchical formation of structures based in cold dark matter 
cosmology \citep[eg.,][]{moo99,bul01}.

In order to quantify how 
important was the accretion mechanism in the formation of the halo, we need 
to find and study the relics of these disrupted galaxies. 
Moreover, their study may reveal a relationship with the population of
dwarf spheroidal (dSph) galaxies that still orbit the Galaxy.
The ultimate goals are to determine the accretion history of the
Galaxy and the fates of its retinue of past and present satellite
galaxies.

We present here the first part of 
a large scale survey of RR Lyrae variable stars (RRs) aimed at
studying the spatial distribution of stars in the halo, 
especially at large distances
from the galactic center. This distant region of the Galaxy
has been poorly studied in the past
mainly because of the lack of wide field CCD cameras able to cover large
regions of the sky, down to magnitudes faint enough to reach the far halo.

\section{The Survey}

Our RR Lyrae survey is being carried at the 1m Schmidt telescope at the 
Observatorio Nacional de Llano del Hato, in Venezuela. This wide-field
telescope is equipped with a large format CCD camera known as the QUEST
camera\footnote{QUEST is an international collaboration (Yale and
Indiana Universities (USA), and Centro de Investigaciones de Astronom{\'\i}a
and Universidad de Los Andes (Venezuela)), whose main goal is to perform
a large scale survey of quasars.} \citep{bal02}. 
The camera is designed to work in
drift-scan mode, a very efficient way to perform surveys. Thus, each night of 
observation we obtain a strip of the sky, which is $2\fdg 3$ wide, and several
hours of right ascension long, in four different filter bandpasses.

In order to maximize the science output of the survey, we were able to combine
several different projects within the same observational scheme. 
Each of these projects required the repeated observation of the same
large region of the sky to the faintest possible limiting magnitude.
Aside from the RR Lyrae survey, some of the projects include: 
searching for quasars through
variability, supernovas \citep{sch01}, transneptunian objects
\citep{fer01} and other solar system science, and
T Tauri stars (Brice\~no et al, this volume).

   \begin{figure}
   \centering
   \includegraphics[width=6.5cm]{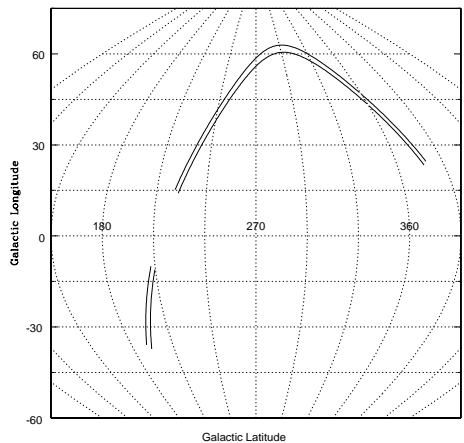}
   \caption{Galactic coordinates of the region of the survey. }
   \label{fig-zone}
   \end{figure}

The first part of the survey was centered at $\delta = -1^\circ$ and covers
a total of 380 deg$^2$ down to a depth of $V\sim 19.7$. It spans a large range
in galactic coordinates (Figure~\ref{fig-zone}), and reveals
RRs lying between 4 and
60 kpc from the Sun. We obtained between 15 and 40 observations along 
each different line of sight.
Due to the very high amount of data ($\sim 1.7$ Terabytes), all the basic 
reduction, astrometry and aperture photometry was done with a custom made
automatic pipeline. 

\begin{figure*}
   \centering
   \resizebox{10.4cm}{!}{\includegraphics{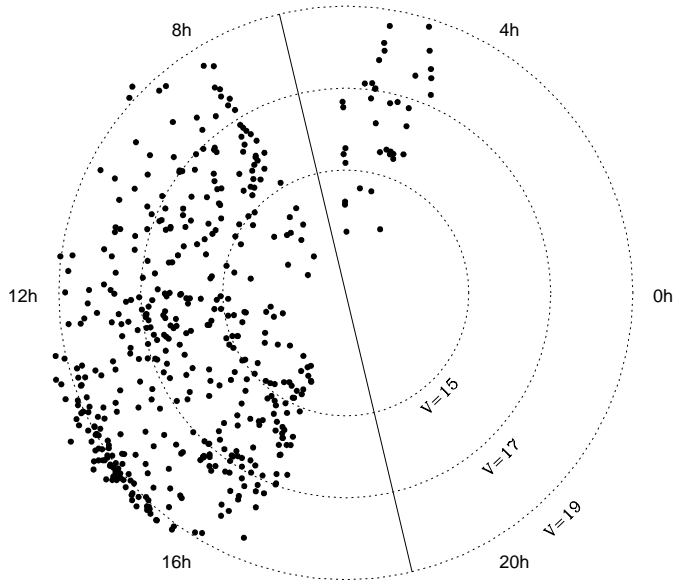}}
   \caption{Radial plot of the distribution of the RRs in right
   ascension. The radial axis is the extinction corrected magnitudes, $V_0$.
   All stars are located in a $2\fdg 3$ wide strip centered at declination
   $-1^\circ 10'.8$. The
   circles correspond to a distance from the Sun of 8, 19 and 49 kpc
   respectively. The solid line indicates the position of the galactic plane.}
   \label{fig-sky}
\end{figure*}

We use relative photometry to construct time series of all objects in the
V band. Variable
objects were chosen using a $\chi^2$ test. To isolate RRs we
imposed further constraints on color (V-R$<0.42$), amplitude of variation
($0.2<\Delta V < 1.6$) and period ($0.15<P<0.9$ days). Finally, visual 
inspection of the phased light curves was done to select RRs
of type ab (asymmetric light curves) and type c (sinusoidal light curves).
The completeness of the survey was estimated by extensive simulations
of artificial light curves, and it is high ($>80\%$) for the types ab but
lower ($40-60\%$) for the type c variables. 

\section{Results}

We have found 497 RRs in the surveyed area, $86\%$ of which are 
newly discovered objects. About 150 of these stars lie more than 30 kpc 
from the galactic center, which is a significant fraction of all the
known objects at large galactocentric distances. For calculating
the distances, we assumed
an absolute magnitude for all RRs of $M_V = +0.55$ \citep{dem00},
and extinction corrections from \citet{sch98}.

The spatial distribution of
RRs can be seen in Figure~\ref{fig-sky}.
We calculated the space density of RRs as a function of 
galactocentric distance for different lines of sight, each covering a small
range in galactic coordinates. In this way, we were able to study differences
in the density profiles at different parts of the halo. 
This technique also enables us 
to detect easily over-densities or clumps in the halo \citep{viv02a,viv02b}.

\begin{figure*}
   \centering
   \resizebox{12.5cm}{!}{\includegraphics{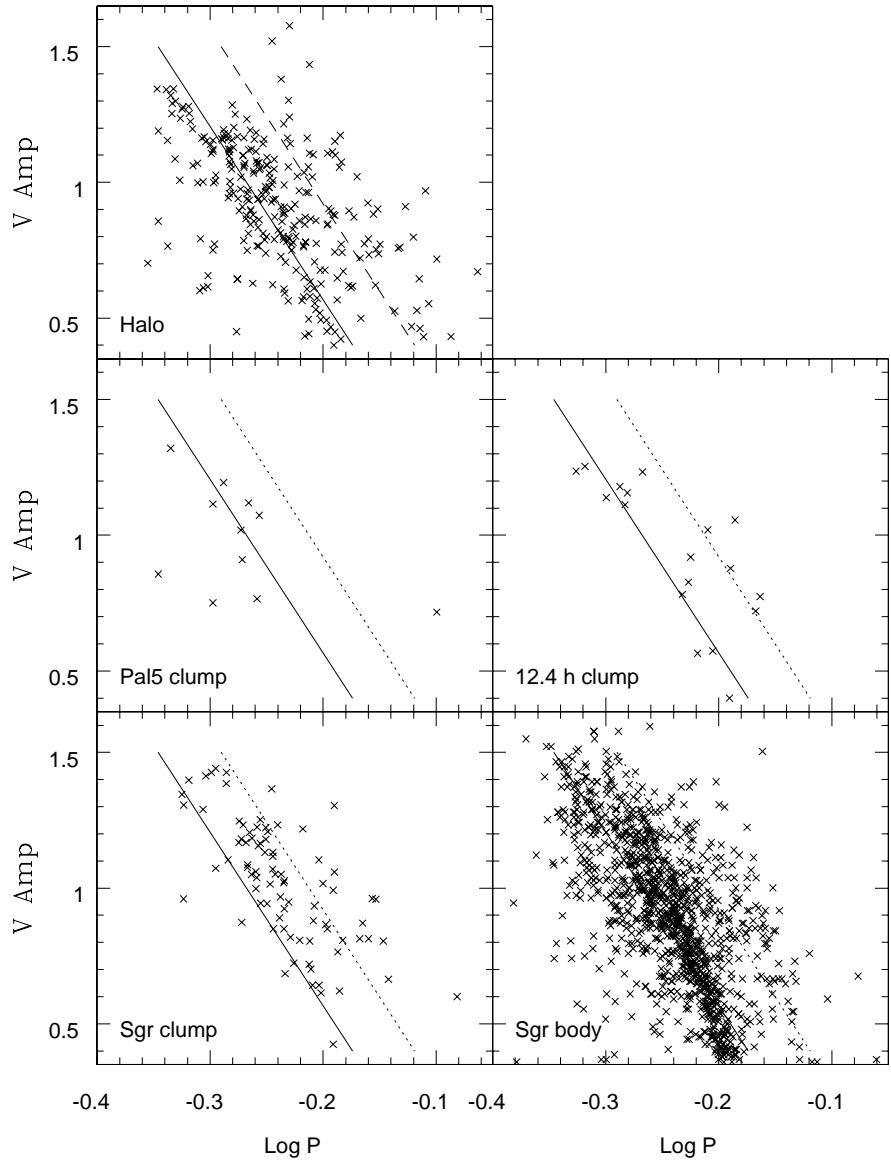}}
   \caption{Period-Amplitude diagrams 
of the major clumps found in the survey. For comparison, we show
a diagram of the general halo RR Lyrae population (which does not include
Sgr stream stars) and a diagram of 1194 RRs in the body of the
Sgr galaxy from the survey by \citet{cse01}. The mean location of RR Lyrae
stars (type ab) in the OoI globular cluster M3, and the
OoII cluster M15, are indicated by the solid and dashed lines respectively.}
   \label{fig-paclumps}
\end{figure*}

\subsection{Shape of the luminous halo}

Since our data spans a large range of both galactic latitude and longitude,
we tested several models of the shape of isodensity contours in the
halo, as a function of distance. 
The model that best fit the data has density contours that are
flattened in the inner halo (galactocentric distance $R_{gal}<20$ kpc), and
spherical at larger distances, as proposed earlier by \citet{pre91}.
However, a halo with density contours slightly flattened (c/a=0.9) 
in the outermost parts cannot be discarded. 

The average space density of RR Lyrae stars between 4 and 60 kpc is well
described by this power law:

$$ \rho (a) = (4.2\pm 1.5) (a/R_0)^{(-3.1 \pm 0.1)} \mbox{kpc}^{-3} $$

\noindent
where $a$ is the semimajor axis of ellipsoids around the galactic center.

There is no sign of a sudden fall off in the space density of RRs
that would indicate the edge of the halo. \citet{ive00} claim that
such an edge could be located at 50-60 kpc, which is right at the
faint limit of our survey.

\subsection{Clumpiness in the halo}

The spatial distribution of RRs in our survey is clearly
clumpy (see Fig.~\ref{fig-sky}). 
In the range $R_{gal} = 20 - 50$ kpc, the variation in the number
of RRs from one part of the sky to another is larger than expected
from pure statistical fluctuations. Although the survey covers an area less 
than $1\%$ of the entire halo, we have detected three strong sub-structures 
in the halo, which are described below. Figure~\ref{fig-paclumps} shows the 
Period-Amplitude diagrams for the RRs (type ab) in these three clumps.

\paragraph{The Sgr stream clump:} is the largest clump found in the survey,
which is probably a consequence of it being part of the present-day
disruption of a dSph galaxy 
by the tidal forces of the Milky Way.
The clump is part of the tidal debris of the Sagittarius (Sgr) dSph 
and it is located $\sim60^\circ$
away from the body of the galaxy. The 84 RRs in the clump 
have a mean magnitude of $\langle V_0 \rangle = 19.1 \pm 0.2$. The clump
is located at 48 kpc from the galactic center and spans an area of about 78 
deg$^2$, at $13\fh 0 < \alpha < 15\fh 4$. The depth along the line of sight is
small, only $\sim 4$ kpc \citep{viv01}.
The presence of this structure has been detected by several surveys with
different tracers \citep{new02,yan00,ive00,iba01}.
The mean period of the RR Lyrae stars in the clump, 0.58 days,
is similar to the value found in the central part of Sgr \citep{cse01}. 

\paragraph{The clump at {\boldmath $\alpha=12\fh 4$}:} this clump 
of 21 RRs, is located only
19 kpc from the Sun ($\langle V_0 \rangle = 16.9 \pm 0.2$). 
It also has high statistical significance (a $5\sigma$ overdensity
compared to the smooth halo contours).
This structure has also been detected as an excess of F stars (halo
turnoff candidates) by \citet{new02} from the {\it Sloan Digital Sky
Survey} (SDSS).
The clump does not seem to be related
with any known dSph galaxy or globular cluster. The mean period and
period-amplitude distribution indicate that 
the RRs do not fit within one Oosterhoff group (Oo), 
which suggests that the clump may be the remnants of an ancient
galaxy rather than a disrupted globular
cluster. 
The radial velocities and metallicities of these stars are being measured, 
and they should shed more light on the origin of this clump.

\paragraph{The Pal 5 clump:} the origin of this clump of 12 stars appears 
to be different from the other two. 
It is located in the same direction ($\alpha=15\fh 3$) and
at the same distance as the globular cluster Pal 5
($\langle V_0 \rangle = 17.1 \pm 0.2$).
The properties of the RRs in the clump suggest an OoI classification, as is the
cluster itself \citep{cle01}. 
A rough estimate of the metallicity of the group based
on the mean period of the RRs is also consistent with measurements of
cluster stars. 
Consequently, this group of RRs, which extends far beyond the tidal
radius of the cluster, have all the right properties to be members of
the cluster.
The recent investigation by \citep{ode01} has detected
tidal tails emanating from Pal 5.
However, our clump stars extend even farther away from the cluster
and over a wider range of directions
than the narrow tidal tails detected by \citeauthor{ode01}

\section{Conclusions}

The streams and sub-structures found in the distribution of RRs in the
outer halo of the Milky Way
resemble the signatures of merger events.
The outer halo is significantly clumpy and this is
consistent with the idea
that accretion of satellites played a major role in the formation
of the outer halo.

Satellite dSph galaxies are the main suspect for being the building blocks of
the halo. If so, the stellar population of the halo should resemble
a composite of several
dSph galaxies.
The pulsational properties of the halo and these galaxies are indeed
similar in that both have wide period-amplitude distributions and mean
periods intermediate
between Oo groups I and II. Although this may not constitute a
strong proof that the halo RR Lyrae population came indeed from disrupted
dSph galaxies, it is at least consistent with this hypothesis.

Our data have confirmed previous studies that showed that the inner
halo is flattened towards the galactic plane.  This may be evidence
that the inner halo formed differently, perhaps as a dissipative
collapse of a protogalactic cloud.

The search for streams in the halo by
QUEST and other surveys like the SDSS 
still cover a small fraction of the sky.
Several streams have been found
but they are still few to allow a description of
the accretion history of the Milky Way. We are continuing the RR
Lyrae survey with the QUEST camera in a second declination strip.
An important by-product of this survey will be a very large database
of variable objects which will be made available to the astronomical
community in the near future. 

\begin{acknowledgements}
The Llano del Hato Observatory is operated by CIDA for the Fondo
Nacional de Investigaciones Cient{\'\i}ficas y Tecnol\'ogicas and the
Ministerio de Ciencia y Tecnolog{\'\i}a of Venezuela.
This research was partially
supported by the National Science Foundation under grant AST-0098428.
We thank all members of the QUEST collaboration for their important 
contribution to this work.
\end{acknowledgements}

\bibliographystyle{aa}

\begin{thebibliography}{}

\bibitem[Baltay et al.(2002)]{bal02} Baltay, C. et al. 2002, \pasp, 114, 780

\bibitem[Bullock, Kravtsov \& Weinberg(2001)]{bul01} Bullock, J. S.,
Kravtsov, A. V. \& Weinberg, D. H. 2001, \apj, 548, 33

\bibitem[Clement et al.(2001)]{cle01} Clement, C. M. et al. 2001, \aj, 122,2587

\bibitem[Cseresnjes(2001)]{cse01} Cseresnjes, P. 2001, \aap, 375, 909

\bibitem[Demarque et al.(2000)]{dem00} Demarque, P., Zinn R., Lee, Y. \&
Yi S. 2000, \aj, 119, 1398

\bibitem[Ferr{\'\i}n et al.(2001)]{fer01} Ferr{\'\i}n, I. et al. 2001,
\apjl, 548, L243

\bibitem[Ibata et al.(2001)]{iba01}  Ibata, R., Lewis, G., Irwin,
M., Totten, E. \& Quinn, T. 2001, \apj, 551, 294

\bibitem[Ivezic et al.(2000)]{ive00} Ivezic, Z. et al. 2000, \aj, 120, 9631

\bibitem[Moore et al.(1999)]{moo99} Moore, B. et al. 1999, \apj, 524, L19

\bibitem[Newberg et al.(2002)]{new02} Newberg, H. J. et al. 2002, \apj, 569,
245

\bibitem[Odenkirchen et al.(2001)]{ode01} Odenkirchen, M. et al. 2001,
\apjl, 548, L165

\bibitem[Preston, Schectman \& Beers(1991)]{pre91} Preston, G. W.,
Schectman, S. A. \& Beers, T. C. 1991, \apj, 375, 121

\bibitem[Schaefer et al.(2001)]{sch01} Schaefer, B. et al. 2001, IAU Circ.
7608

\bibitem[Schlegel, Finkbeiner \& Davis(1998)]{sch98} Schlegel, D. J.,
Finkbeiner, D. P. \& Davis, M. 1998, \apj, 500, 525

\bibitem[Vivas et al.(2001)]{viv01} Vivas, A. K. et al. 2001, \apjl, 554, L33

\bibitem[Vivas \& Zinn(2002)]{viv02a} Vivas, A. K. \& Zinn, R. 2002, in
``The Shape of Galaxies and Their Dark Halos'' P. Natarajan (ed). 
World Scientific, 210

\bibitem[Vivas(2002)]{viv02b} Vivas, A. K. 2002, PhD Thesis, Yale University

\bibitem[Yanny et al.(2000)]{yan00} Yanny, B. et al. 2000, \apj, 540, 825

\end{thebibliography}

\end{document}